\documentclass{article}  
\usepackage{amsmath}
\usepackage{amssymb}
\usepackage{citesort}
\begin{document}

\title{Bogomolny equation for the BPS Skyrme model from the strong necessary conditions}
\author{{\L}. T. St\c{e}pie\'{n}\footnote{The Pedagogical University of Cracow, ul. Podchora\c{}\.{z}ych 2, 30-084 Krak\'{o}w, Poland, \ 
sfstepie@cyf-kr.edu.pl, lstepien@up.krakow.pl}}

\maketitle

\begin{abstract}
We present a systematic tool of derivation of the Bogomolny equation for the BPS Skyrme model. Furthermore, we find a generalization of the Bogomolny equation to the case corresponding with a non-zero value of the external pressure. The method is based on the concept of strong necessary conditions and can be applied to any Skyrme like theory. 
\end{abstract}



\section{Introduction}

The Skyrme model \cite{skyrme} is one of the most popular effective theories of the low energy QCD describing non-perturbative excitations i.e., baryons and atomic nuclei as coherent solutions in the "mesonic fluid", that is (semiclassically quantized) solitons in a theory with pionic degrees of freedom \cite{nappi}, \cite{wood}, \cite{manton} (fields in the Lagrangian).   
The original version of this model is given by the Lagrangian consisting of two terms: 1) $\mathcal{L}_{2}$, which is the sigma-model term, representing here the kinetic energy of pions and 2) $\mathcal{L}_{4}$ - the Skyrme-term, necessary for overcoming the consequences of Derrick-Hobart theorem
   \begin{equation}
   \mathcal{L} = \mathcal{L}_{2} + \mathcal{L}_{4},
   \end{equation}
    where 
   \begin{eqnarray}
   \mathcal{L}_{2} &=& -\frac{f^{2}_{\pi}}{4} Tr(U^{\dagger} \partial_{\mu} U U^{\dagger} \partial^{\mu} U),\\
   \mathcal{L}_{4} &=& \frac{1}{32 e^2} Tr([U^{\dagger} \partial_{\mu} U, U^{\dagger} \partial_{\nu} U]^{2}).
   \end{eqnarray}
For recent phenomenological results see f.g. \cite{lau}.

Since the Skyrme model is an effective theory of the strong interactions it is possible (and perhaps unavoidable) to add more (in practice infinitely many) terms. However, if we impose natural conditions that the model should be Lorentz invariant and possess a usual Hamiltonian description then only two additional terms can be included. Namely a non-derivative part i.e., potential $\mathcal{L}_0$ and a sextic term
 \begin{equation}
 \mathcal{L}_{6} = -\lambda^{2} \pi^{4} \mathcal{B}^{2}_{\mu},
 \end{equation}
where $\mathcal{B}_{\mu}$ is the topological current:
\begin{equation}
  \mathcal{B}_{\mu} = \frac{1}{24 \pi^{2}} Tr(\varepsilon^{\mu \nu \rho \sigma} U^{\dagger} \partial_{\nu} U U^{\dagger} 
  \partial_{\rho} U U^{\dagger} \partial_{\sigma} U)
  \end{equation} 
and $\int d^{3}x \mathcal{B}_{0}$, one identifies with baryon number.

It is a matter of fact that such a general Skyrme model
\begin{equation}
  \mathcal{L} = \mathcal{L}_0+\mathcal{L}_{2} + \mathcal{L}_{4}+\mathcal{L}_6,
\end{equation}
has a BPS limit. Indeed, if one choses the pertinent constant such that the sigma model part and the Skyrme term vanish then we reach the BPS Skyrme model \cite{BPS} (see also \cite{Marl})
\begin{equation}
  \mathcal{L}_{BPS} = \mathcal{L}_0+\mathcal{L}_6.
\end{equation}
This limit is not only of purely mathematic interest but it corresponds to a very well motivated 
idealization of nuclear matter. Two most pronounced properties of atomic nuclei and nuclear matter are: very small values of the binding energies and fluid property of nuclear matter (compare liquid droplet model). The BPS Skyrme model realizes these two crucial features. It is a BPS theory which means that classically solitons (atomic nuclei) have zero binding energies \cite{BPS} (physical binding energies for the most abundant nuclei can be obtained by the semiclassical quantization and inclusion of the Coulomb energy together with the isospin breaking \cite{bind}). Furthermore, the model is exactly a field theoretical description (in Eulerian formulation) of a perfect fluid \cite{chem-pot}. Obviously, in the full effective action one cannot neglect the $\mathcal{L}_2$ and $\mathcal{L}_4$ terms but the they should provide rather small contribution to the total soliton mass if one wants to keep the binding energies in an agreement with the experimental data \cite{bind}, \cite{Sp2}, \cite{Sp}. Therefore, the BPS Skyrme model seems to give leading contribution to some bulk observables of atomic nuclei and nuclear matter. In a consequence, one can start studying such quantities in the BPS limit of the full Skyrme action. This is interesting observation as the BPS Skyrme model is a {\it solvable} solitonic model, where many results can be found in an exact analytical way. This for example has recently led to a derivation of exact gravitating BPS skyrmions with topological charge $B \sim 10^{57}$ interpreted as neutron stars, where the full back reaction of the gravity on the matter was taken into account \cite{stars1}, \cite{stars2} (gravitating BPS skyrmions has been also analyzed in \cite{gud}). 

The Bogomolny equation for the BPS Skyrme model can be derived as follows \cite{BPS}, \cite{Sp1}. Here $\mathcal{L}_0=-\mu^2V$ ($\mu$ is a constant) and we use the usual parametrization of the Skyrme field $U$,
  \begin{equation}
  U = e^{i \chi \vec{n} \cdot \vec{\sigma}},
  \end{equation}
  where $\vec{\sigma}$ are the Pauli matrices and the unit vector field $\vec{n}$ is related to the field 
  $\omega \in \mathbb{C}$, by the stereographic projection
  \begin{equation}
  \vec{n} = \bigg[\frac{\omega + \omega^{\ast}}{1+ \omega \omega^{\ast}}, -i \frac{\omega - \omega^{\ast}}{1+ \omega\omega^{\ast}}, 
  \frac{1 - \omega \omega^{\ast}}{1+ \omega \omega^{\ast}}\bigg], \  \mid \vec{n} \mid^{2} = 1. 
  \end{equation} 
 Hence, the Lagrangian has the form,
  \begin{equation}
  \mathcal{L} = -\frac{\lambda^{2} \sin^{4}{\chi}}{(1+\omega\omega^{\ast})^{4}} (\varepsilon^{\mu\nu\rho\sigma}\chi_{,\nu} \omega_{,\rho} \omega^{\ast}_{,\sigma})^{2} - \mu^{2} V,
  \end{equation}
where $\chi_{,\nu} = \frac{\partial \chi}{\partial x^{\nu}}$ etc. and the potential is a function of $\chi$. The BPS bound (Bogomolny-Prasad-Sommerfield) for this model, has the form, 
   \begin{gather}
      E = \int d^{3} x \bigg(\frac{\lambda^{2} \sin^{4}{\chi}}{(1+\omega\omega^{\ast})^{4}} (\varepsilon^{kmn} i \chi_{,k} \omega_{,m} \omega^{\ast}_{,n})^{2} + \mu^{2} V(\chi)\bigg)  \geq 2\lambda \mu \pi^{2} \langle \sqrt{V} \rangle_{S^{3}} \mid B \mid. 
   \end{gather}
This bound is saturated and it yields the BPS equation \cite{BPS}, \cite{Sp1}.

  \begin{equation}
  \frac{\lambda \sin^{2}{\chi}}{(1+\omega\omega^{\ast})^{2}} \varepsilon^{kmn} i \chi_{,k} \omega_{,m} \omega^{\ast}_{,n} = \mp \mu \sqrt{V}.
  \end{equation} 
  Let us remark that, on contrary to the Bogomolny bound for the BPS Skyrme model, the Bogomolny bound for the usual Skyrme model, is not saturated \cite{MS}. However, in the paper \cite{Canf_etal1} (based on the approaches developed in \cite{Canf_etal2}), some novel BPS bound for the usual four dimensional Skyrme model, was obtained, and the saturation of this bound was showed.  

In this paper, we derive the Bogomolny equation for BPS family Skyrme model with a general form of the potential $V=V(\chi, \omega, \omega^{\ast})$, in a systematic way, by applying the concept of strong necessary conditions, which was introduced firstly in \cite{S1} and developed in  \cite{SWL1} and in \cite{SWL2}. The derivation of Bogomolny equations for some other field theory systems in lower dimension, by using this method was presented in \cite{SSS}, \cite{S},  \cite{SSS2}. It was also applied to baby Skyrme theories and their gauged version, in \cite{LS1} and \cite{LS2}, correspondingly. The main motivation, besides testing the method in 3 dimensional solitonic system, is a searching for a universal tool which would allow for derivation of Bogomolny equations for the BPS Skyrme model after its coupling to electrodynamics and/or gravity. 

 This paper is organized, as follows. The next section contains a short description of the concept of strong necessary conditions. In the section 3, we derive Bogomolny equation for BPS family Skyrme model, by using this concept. In the section 4, we sum up the results, obtained in this paper.

   \section{A short presentation of the concept of strong necessary conditions}

 Instead of considering of the Euler-Lagrange equations, 
 
 \begin{equation}
 F_{,u} - \frac{d}{dx}F_{,u_{,x}} - \frac{d}{dy}F_{,u_{,y}}=0, \label{el}
 \end{equation}
  following from the varying of the functional
 
 \begin{equation}
 \Phi[u]=\int_{E^{2}} F(u,u_{,x},u_{,y}) \hspace{0.05 in} dxdy, \label{functional}
 \end{equation}
 we take into account, strong necessary conditions, \cite{S1}, \cite{SWL1}, \cite{SWL2}

 \begin{gather}
   F_{,u}=0, \label{silne1} \\
   F_{,u_{,x}}=0, \label{silne2} \\
   F_{,u_{,y}}=0, \label{silne3}
 \end{gather} 
 where $F_{,u} \equiv \frac{\partial F}{\partial u}$, etc.
   
 It is obvious that all solutions of the system of the equations (\ref{silne1}) - (\ref{silne3}) satisfy the Euler-Lagrange equation (\ref{el}). However, if these solutions exist, they are very often trivial solutions. So, in order to extend the set of these solutions, we make gauge transformation of the functional (\ref{functional})
 
  \begin{equation}
  \Phi \rightarrow \Phi + Inv, \label{gauge_transf}
  \end{equation}
  where $Inv$ is such functional that its local variation vanishes, with respect to $u(x,y)$:
 $\delta Inv \equiv 0$.
 
  Owing to this feature, the Euler-Lagrange equations (\ref{el}) and the Euler-Lagrange equations resulting from  requiring of the extremum of $\Phi + Inv$, are equivalent.
 On the other hand, the strong necessary conditions (\ref{silne1}) - (\ref{silne3}) are not invariant with respect to the gauge transformation (\ref{gauge_transf}). Hence, we may expect to obtain non-trivial solutions. Let us note that the order of the system  of the partial differential equations, consituted by strong necessary conditions (\ref{silne1}) - (\ref{silne3}), is less than the order of Euler-Lagrange equations (\ref{el}). 

 \section{The derivation of the BPS equation for the BPS Skyrme model}

Let us now show how this framework works in practice and derive the Bogomolny equation for the BPS Skyrme model.

 According to the idea presented in the previous section, we make the gauge transformation of the density of the energy functional of the BPS Skyrme model: $\mathcal{H} \longrightarrow \mathcal{\tilde{H}}$, however, in order to generalize the investigations, we consider
  \begin{gather}
   H = \int d^{3} x \mathcal{H} = \int d^{3} x (f \cdot (\varepsilon^{kmn} i \chi_{,k} \omega_{,m} \omega^{\ast}_{,n})^{2} + V),  \label{hamiltonian}
  \end{gather}
   where $f, V \in \mathcal{C}^{1}$ are some unspecified functions of $\chi, \omega, \omega^{\ast}$. (For the sake of generality we leave $f$ as a completely unspecified function which correspond to a general target space geometry. The usual $S^3$ target space is obtained by assumption that $f=\frac{\sin^4 \chi}{(1+|\omega|^2)^4}$.) Next, we have to establish the general form of the density of the topological invariant $I_{1}=\int d^{3}x \mathcal{I}_{1}$, i.e. $\delta I_{1} \equiv 0$. It turns out that this 
  density has the following form:
  \begin{gather}
  \mathcal{I}_{1} = G_{1} \varepsilon^{kmn}\chi_{,k} \omega_{,m} \omega^{\ast}_{,n}, \label{glowny_niezm}
  \end{gather}
where $G_{1} \in \mathcal{C}^{1}$ is some arbitrary function of $\chi, \omega,\omega^{\ast}$.
\\
The density of the gauged energy functional has the following form:
  \begin{gather}
  \mathcal{\tilde{H}} = f \cdot (\varepsilon^{kmn} i \chi_{,k} \omega_{,m} \omega^{\ast}_{,n})^{2} + V + \sum^{4}_{l=1} \mathcal{I}_{l},
  \end{gather}
where  $f=f(\chi, \omega, \omega^{\ast}), V = V(\chi, \omega, \omega^{\ast}) \in \mathcal{C}^{1}$. $\mathcal{I}_{1}$ is given by (\ref{glowny_niezm}) and $\mathcal{I}_{p+1}$ are the densities of so-called divergent invariants: $\mathcal{I}_{p+1} = D_{p}G_{p+1}, 
 p=1,...,3$. $G_{1}$ and $G_{p+1} \in \mathcal{C}^{1}$, are some functions of $\chi, \omega, \omega^{\ast}$, to be determined later.\\
 The applying of strong necessary conditions, gives the so-called dual equations:
  \begin{gather}
  \mathcal{\tilde{H}}_{,\chi} : f_{,\chi} \cdot  (\varepsilon^{kmn} i \chi_{,k} \omega_{,m} \omega^{\ast}_{,n})^{2} + V_{,\chi} +  
  G_{1,\chi} \varepsilon^{kmn}\chi_{,k} \omega_{,m} \omega^{\ast}_{,n} + \sum^{3}_{p=1} D_{p} G_{p+1,\chi} = 0, \label{gorne1} \\
  \mathcal{\tilde{H}}_{,\omega} : f_{,\omega} \cdot  (\varepsilon^{kmn} i \chi_{,k} \omega_{,m} \omega^{\ast}_{,n})^{2} + V_{,\omega} +  
  G_{1,\omega} \varepsilon^{kmn}\chi_{,k} \omega_{,m} \omega^{\ast}_{,n} + \sum^{3}_{p=1} D_{p} G_{p+1,\omega} = 0, \label{gorne2} \\
  \mathcal{\tilde{H}}_{,\omega^{\ast}} : f_{,\omega^{\ast}} \cdot  (\varepsilon^{kmn} i \chi_{,k} \omega_{,m} \omega^{\ast}_{,n})^{2} + V_{,\omega^{\ast}} +  
  G_{1,\omega^{\ast}} \varepsilon^{kmn}\chi_{,k} \omega_{,m} \omega^{\ast}_{,n} + \sum^{3}_{p=1} D_{p} G_{p+1,\omega^{\ast}} = 0, \label{gorne3} \\
  \mathcal{\tilde{H}}_{,\chi_{,r}} : 2 f \cdot \varepsilon^{rsj} i \ \omega_{,s} \omega^{\ast}_{,j}  (\varepsilon^{kmn} i \chi_{,k} \omega_{,m}  
  \omega^{\ast}_{,n}) + G_{1} \varepsilon^{rsj} \omega_{,s} \omega^{\ast}_{,j} + G_{r+1,\chi} = 0 , \label{dolne1}\\
  \mathcal{\tilde{H}}_{,\omega_{,r}} : 2 f \cdot \varepsilon^{srj} i \ \chi_{,s} \omega^{\ast}_{,j}  (\varepsilon^{kmn} i \chi_{,k} \omega_{,m}  
  \omega^{\ast}_{,n}) + G_{1} \varepsilon^{srj} \chi_{,s} \omega^{\ast}_{,j} + G_{r+1,\omega} = 0, \label{dolne2}\\
  \mathcal{\tilde{H}}_{,\omega^{\ast}_{,r}} : 2 f \cdot \varepsilon^{sjr} i \ \chi_{,s} \omega_{,j}  (\varepsilon^{kmn} i \chi_{,k} \omega_{,m}  
  \omega^{\ast}_{,n}) + G_{1} \varepsilon^{sjr} \chi_{,s} \omega_{,j} + G_{r+1,\omega^{\ast}} = 0. \label{dolne3}
  \end{gather} 
Now we need to make the equations (\ref{gorne1}) - (\ref{dolne3}), self-consistent. At first, we want  the equations (\ref{dolne1}) - (\ref{dolne3}) are self-consistent.
  In this order, we put

   \begin{equation}
   \begin{gathered}
   2 i f \cdot  (\varepsilon^{kmn} i \chi_{,k} \omega_{,m} \omega^{\ast}_{,n}) + G_{1} = 0, \label{uzg} \\
   G_{r+1}=const, r=1,2,3.
   \end{gathered}
   \end{equation}    
 Hence, instead of three equations (\ref{dolne1}) - (\ref{dolne3}), we have one equation:
   \begin{gather}
   2 i f \cdot  (\varepsilon^{kmn} i \chi_{,k} \omega_{,m} \omega^{\ast}_{,n}) + G_{1} = 0,  \label{Bogomolny}
   \end{gather}
where $f=f(\chi, \omega, \omega^{\ast}) \in \mathcal{C}^{1}, G_{1}=G_{1}(\chi, \omega, \omega^{\ast}) \in \mathcal{C}^{1}$. Next, we eliminate from 
   (\ref{gorne1}) - (\ref{gorne3}), all terms including the derivatives of the fields, by using (\ref{uzg}). Hence, we have three equations for the potential $V(\chi, \omega, \omega^{\ast})$

   \begin{equation}
   \begin{gathered}
   -\frac{1}{4} \frac{f_{,\chi} G^{2}_{1}}{f^{2}} +  V_{,\chi} + \frac{1}{2}  \frac{G_{1,\chi} G_{1}}{f} = 0,\\
   -\frac{1}{4} \frac{f_{,\omega} G^{2}_{1}}{f^{2}} + V_{,\omega} + \frac{1}{2}  \frac{G_{1,\omega} G_{1}}{f} = 0,\\
   -\frac{1}{4} \frac{f_{,\omega^{\ast}} G^{2}_{1}}{f^{2}} + V_{,\omega^{\ast}} + \frac{1}{2}  \frac{G_{1,\omega^{\ast}} G_{1}}{f} = 0.\\
   \end{gathered}
   \end{equation}
Their solution is

   \begin{gather}
   V(\chi, \omega, \omega^{\ast}) = -\frac{1}{4} \frac{G^{2}_{1} - f c_{1}}{f},  \   c_{1} = const. \label{potencjal}
   \end{gather}
   Such a relation between $V, f $ and $G_1$ allows us to remove $G_1$ from (\ref{Bogomolny}). Then after putting $c_{1} = - c_{2}$, we find  
   \begin{equation}
      2 \sqrt{f} \cdot  (\varepsilon^{kmn} i \chi_{,k} \omega_{,m} \omega^{\ast}_{,n}) = \mp \sqrt{4V+c_{2}}, 
   \end{equation} 
For $c_{2}=0$ we re-derive the Bogomolny equaiton for the BPS Skyrme model (where function $f$ is chosen in the proper way). For $c_{2} >0$ we find non-Bogomolny equation which, in fact, coincides with non-zero pressure equation for the BPS Skyrme model \cite{term}. Therefore, the constant found in our construction can be related with the pressure. This is an interesting, unexpected observation that our method not only leads to the Bogomolny equation but also includes the non-zero pressure generalization.

\section{Summary}

In this paper we have applied the concept of strong necessary conditions method to the BPS Skyrme model. We have derived the Bogomolny equation for the BPS family Skyrme model, with the general form of the potential $V=V(\chi, \omega, \omega^{\ast})$ as well as its non-zero pressure generalization. This shows that our method works also for such 3+1 dimensional solitonic theory. 

There are several directions in which the present work can be continued. 

First of all, one can couple the BPS Skyrme model to the Maxwell field in the standard minimal way, as it was done for the baby Skyrme model \cite{schr1}. In 2+1 dimension the BPS baby  model has (saturated) Bogomolny bound \cite{Sp1}, \cite{LS1}, and the analogical result was also obtained for the gauged 2+1 dimensional BPS baby Skyrme model in \cite{LS2}, \cite{BPS-g}. Although it is rather not expected to find Bogomolny equations for 3+1 dimensional gauged BPS Skyrme model in a generic Maxwell field case, one can hope that it can happen for some special forms of magnetic field. This can be of some importance if one wants to use the BPS Skyrme model for description of magnetic neutron stars \cite{He}. 

Secondly, one can apply the method for gravitating skyrmions. Then again no saturated Bogomolny equations are expected to exist, but one can try to find an analytical insight into the problem of gravitational mass loss (which in a sense describes a departure from the original BPS property of the non-gravitating model), which is an important observable for neutron stars. 

Finally, there have been recently proposed several methods for derivation of Bogomolny equations in some specific field theoretical models \cite{bazeia}, \cite{bogom}. It would be nice to understand possible relations between them and this one, applied in this paper.

\section{Acknowledgments}
The author thanks to Prof. F. Canfora, for pointing out the references \cite{Canf_etal1} and  \cite{Canf_etal2}.

\section{Computational resources} 
  
   This research was supported by PL-Grid Infrastructure.

\section*{References}

\end{document}